\documentclass[12pt,preprint]{aastex}

\usepackage{graphics}
\usepackage{graphicx}
\usepackage{multicol} 

\let\oldbibliography\thebibliography
\renewcommand{\thebibliography}[1]{%
  \oldbibliography{#1}%
  \setlength{\parskip}{0.0pt}%
  \setlength{\itemsep}{0.0pt}%
  }

\ifx\pdfoutput\undefined
\usepackage{graphicx}
\else
\usepackage{epstopdf}
\fi

\usepackage[empty]{fullpage} 
\usepackage[small,compact,sf,bf]{titlesec}

\usepackage{times}           

\begin{document}

\newcommand \Webb {\textit{James Webb Space Telescope}}
\newcommand \JWST {JWST}

\newcommand \cf {cf.}
\newcommand \eg {{e.g., }}
\newcommand \ie {{i.e.,}}
\newcommand \viz {{viz.,}}
\newcommand \lap {$\stackrel{<}{\sim}$}
\newcommand \gap {$\stackrel{>}{\sim}$}

\title{Planetary system, star formation, and black hole science with  \\
	non-redundant masking on space telescopes}

\author{ Anand Sivaramakrishnan\altaffilmark{1} (AMNH), Peter Tuthill (U. Sydney), Frantz Martinache (Subaru), \\
Michael Ireland (U. Sydney), James Lloyd (Cornell), Marshall Perrin (UCLA),  \\
R\'emi Soummer (STScI), Barry McKernan and Saavik Ford (CUNY)
\altaffiltext{1}{anand@amnh.org \hfill White Paper in NAS ASTRO2010 Decadal Survey, 31 March 2009} }
%
\begin{abstract} 
Non-redundant masking (NRM) is a high contrast, high resolution
technique relevant to future space missions concerned with extrasolar
planetary system and star formation, as well as general high angular
resolution galactic and extragalactic astronomy.
NRM enables the highest angular resolution science
possible given the telescope's diameter and operating wavelength. It also
provides precise information on a telescope's optical state.
NRM relies on its high quality self-calibration
properties and the robustness of interferometric techniques, whereas
coronagraphy requires exquisite wavefront quality. 
Stability \textit{during} an observation sets fundamental NRM contrast limits.
A non-redundant mask was recently added to JWST's Fine Guidance
Sensor Tunable Filter Imager (FGS-TFI) instrument,
bringing a no-cost, no-impact boost
in angular resolution that complements JWST's coronagraphs.
The JWST NRM search space lies between 50 and 400 mas
at 3.8 to 5\micron, even if the telescope's image quality does not meet requirements.
JWST's NRM will produce 10 magnitudes of contrast in a 10~ks exposure on an M=7 star,
placing Taurus protoplanets and nearby Jovians younger than 300 Myr
within JWST's reach. 

Future space telescopes can improve vastly on JWST's NRM by utilizing
more refined observing methods, and instrumentation designed to take
full advantage of NRM's high dynamic range.
The ATLAST 16~m 
design can deliver 10 to 12 magnitudes
of contrast between 0.7 to 6 mas at 0.1 \micron.
On an 8~m telescope at 0.1 \micron\ NRM resolution is almost ten times
finer than ALMA's finest resolution.
Polarization with space-based NRM opens new vistas of
astrophysics in planetary system and star formation as well as
AGNs and structure around galactic black hole candidates.
Space NRM explores areas inaccessible to both
JWST coronagraphs and future 30-m class ground-based telescopes. 
Ground-based NRM is limited by atmospheric variability. 

Optimization of space-based NRM requires consideration of 
flat fielding accuracy, target placement
repeatability, charge diffusion, intrapixel sensitivity, image persistence, charge
transfer efficiency, guiding, wavefront stability, pupil wander, and other details.
We must assess NRM contrast limits realistically
to understand the science yield of  NRM in space,
and, simultaneously,
develop NRM science for planet and star formation and extragalactic science in
the UV-NIR, to help steer high resolution space-based astronomy in the coming decade.
\end{abstract}
				\clearpage   

\maketitle

\noindent \textbf{\textsf{\Large \noindent Planetary system, star formation, and black hole science \\
with  	non-redundant masking on space telescopes} } 

\renewcommand{\thepage}{\arabic{page}}

\section{Introduction}
%

\begin{figure}
 \centerline{ \includegraphics[angle=0,scale=0.8]{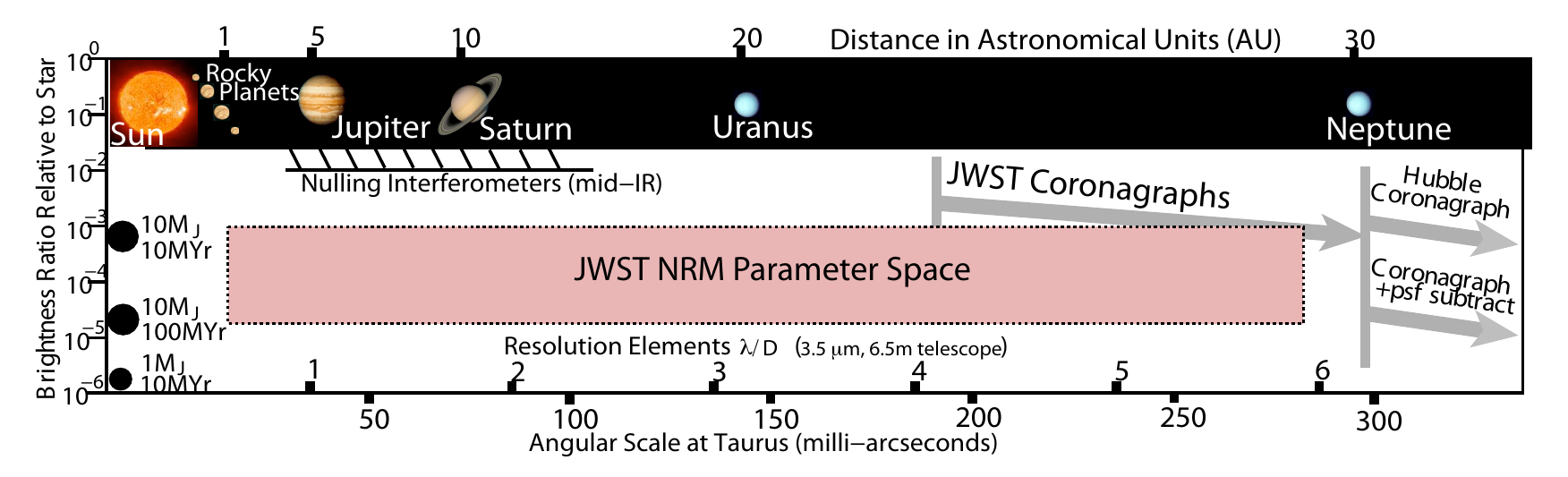} }
 \caption{ JWST FGS Tunable Filter Imager's NRM search space.
}
\label{jamienrmsearchfig}
\end{figure}
Major ground-based observatories have successfully implemented non-redundant aperture-masking
interferometry using existing instruments and observing procedures.
Landmark discoveries such as dusty disks imaged around young stellar objects,
mass-loss shells of evolved stars, low mass companions to nearby stars,
and the fascinating time-varying spiral plumes surrounding dusty Wolf-Rayet systems
have been reported amongst the 50-odd  peer-reviewed papers on NRM science 
\citep{1998SPIE.3350..839T,1999ApJ...525L..97M,1999ApJ...512..351M,2000PASP..112..555T,2001Natur.409.1012T,2002ApJ...577..826T,2005ApJ...624..352T,2006Sci...313..935T,2006ApJ...650L.131L,2006ApJ...649..389P,2007ApJ...661..496M,2007Sci...316..247T,2008ApJ...678L..59I,2008ApJ...678..463I,2008ApJ...675..698T,2008arXiv0809.0895M}.
However, atmospheric transmission and instabilities place significant
limitations on ground-based non-redundant masking (NRM) performance.
This will change with the advent of NRM on board space telescopes.

\begin{figure}
 \centerline{ \includegraphics[angle=0,scale=0.8]{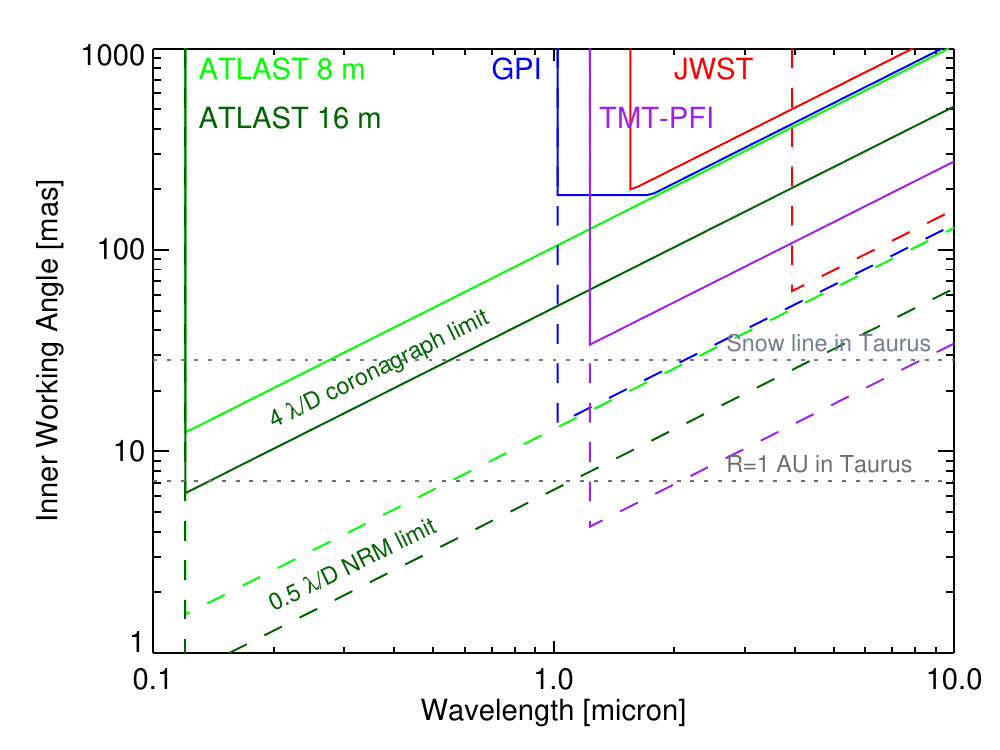} 
\includegraphics[angle=0,scale=0.415]{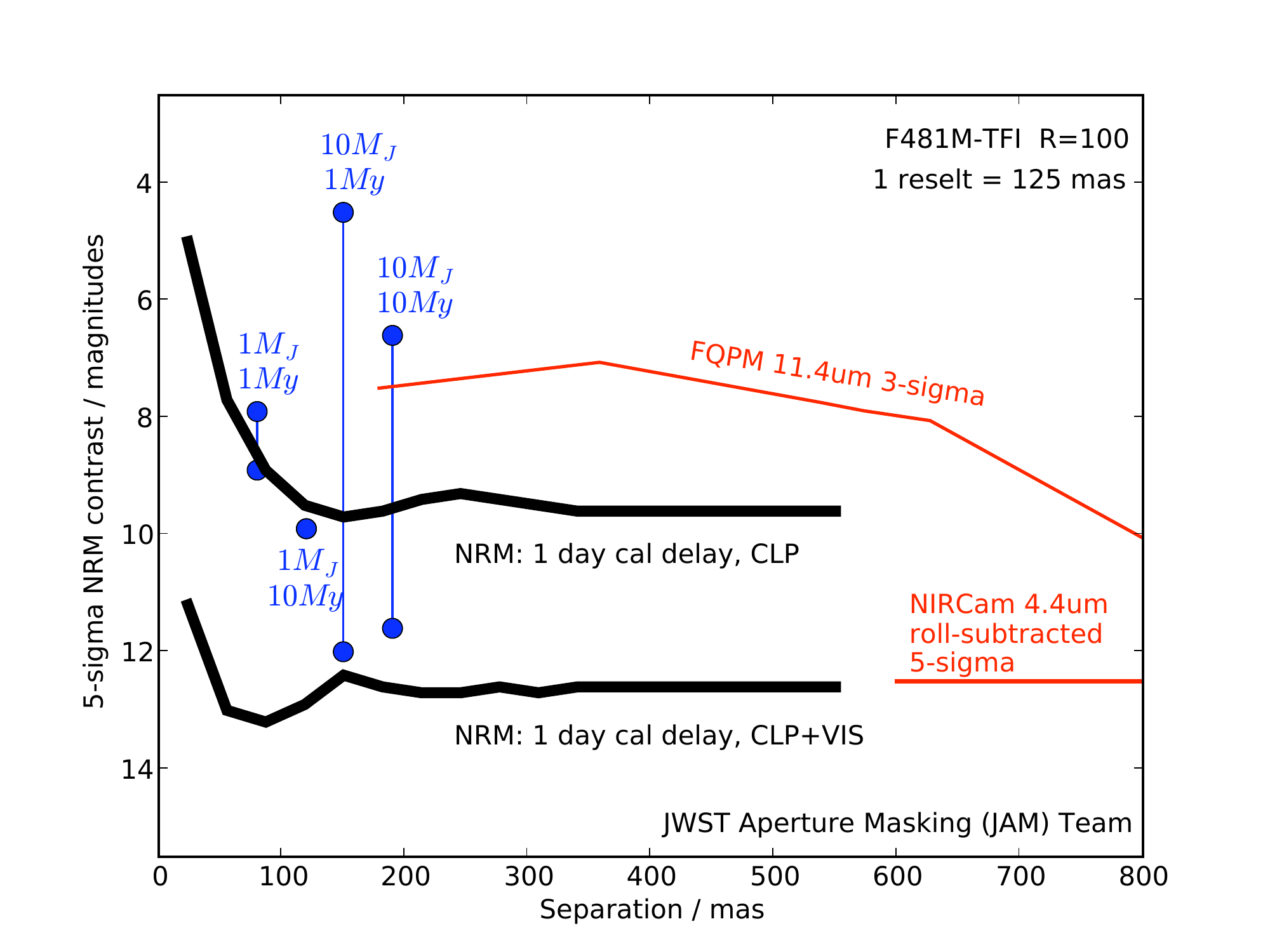} }
 \caption{ NRM and coronagraphy in space.
 \textit{Left:} Solid lines show regions accessible with a $4\lambda/D$ inner working angle (IWA)
 coronagraph (e.g. JWST NIRCam), while dotted lines mark the $0.5
 \lambda/D$ NRM limit.
   \textit{Right:} Estimated dynamic range using closure phase (CLP) as well as
   closure phase and fringe visibility (CLP+VIS) data for 1\% bandpass
   NRM imaging at 4.81 \micron\ using the 7-hole mask (shown in Fig.~\ref{mask7fig})
   in \JWST's Fine Guidance Sensor's Tunable Filter Imager.
   Protoplanets in Taurus will be detectable around an $M=7$ star in a
   10ks exposure, and a one day delay between the target and a
   calibrator star. 
   Photon noise with this exposure limits the dynamic range to 10
   magnitudes. The range of estimated contrasts between a solar type
   star and 1 and 10 Jupiter mass planets, at ages of 1 and 10 Myr
   \citep{2003A&A...402..701B,2007ApJ...655..541M} are shown in blue
   (at arbitrary separations).
   10\% bandwidths will produce similar results with 1ks exposures. 
}
\label{nrmmdpfig}
\end{figure}

Detailed simulations of NRM performance on the  \Webb\  (\JWST) 
(Fig.~\ref{jamienrmsearchfig})
demonstrate the exciting planetary science enabled by the recent addition
of NRM to \JWST's suite of established instruments, emphasizing
the importance of this technique to other future space-based observatories
\citep{JAM09}.
On \JWST\ NRM will widen the telescope's science reach to include a
unique combination of wavelength and angular resolution regimes
inaccessible from the ground, even with the advent of extreme adaptive
optics (ExAO) and coronagraphs behind 30~m extremely large telescopes
of the future.
NRM used at 4 \micron\  will bring  within \JWST's purview
warm extrasolar jovians within 4 to 30~AU  of F, G, and K dwarfs 30~pc
from the Sun (Fig.~\ref{nrmmdpfig}).

NRM adds high resolution capability to large filled-aperture space
telescopes without sacrificing their wide utility for general astrophysical observations.
The NRM search space is complementary to future coronagraphic space missions
dedicated to extrasolar planet imaging and characterization \citep{2003ApJ...582.1147K},
as well as to ground-based ExAO instruments (e.g. \citealt{2006SPIE.6272E..18M}),
in that it yields moderate contrasts between 0.5 and $4\lambda/D$
($\lambda$ being the observing wavelength, $D$ the telescope diameter).
Diffraction-limited stellar coronagraphy, on the other hand, typically covers a
search space at higher contrast, but wider separations.
Contrast-enhancing techniques such as Angular or Spectral Differential Imaging
(\eg\ \citet{2006ApJ...641..556M,2007ApJ...661.1208L} and references therein)
are inefficient at the small angular separations covered by NRM.

\citet{JAM09} examined the scientific merit and feasibility of NRM
on JWST FGS-TFI
between 3.8 \micron\  and 5 \micron, operating at a spectral resolution of 100
set by the \'etalon
(NRM designs can perform just as well in 20\% bandpass filters).
The inner and outer working angles (IWA \& OWA) of NRM scale with wavelength.
CCD detectors in the visible/UV can be calibrated more accurately 
than JWST's IR focal plane arrays, resulting in improved NRM contrast.
%

\section{NRM Interferometry and its limits}

%
 \begin{figure}
 \centerline{
   \includegraphics[angle=0,scale=0.3]{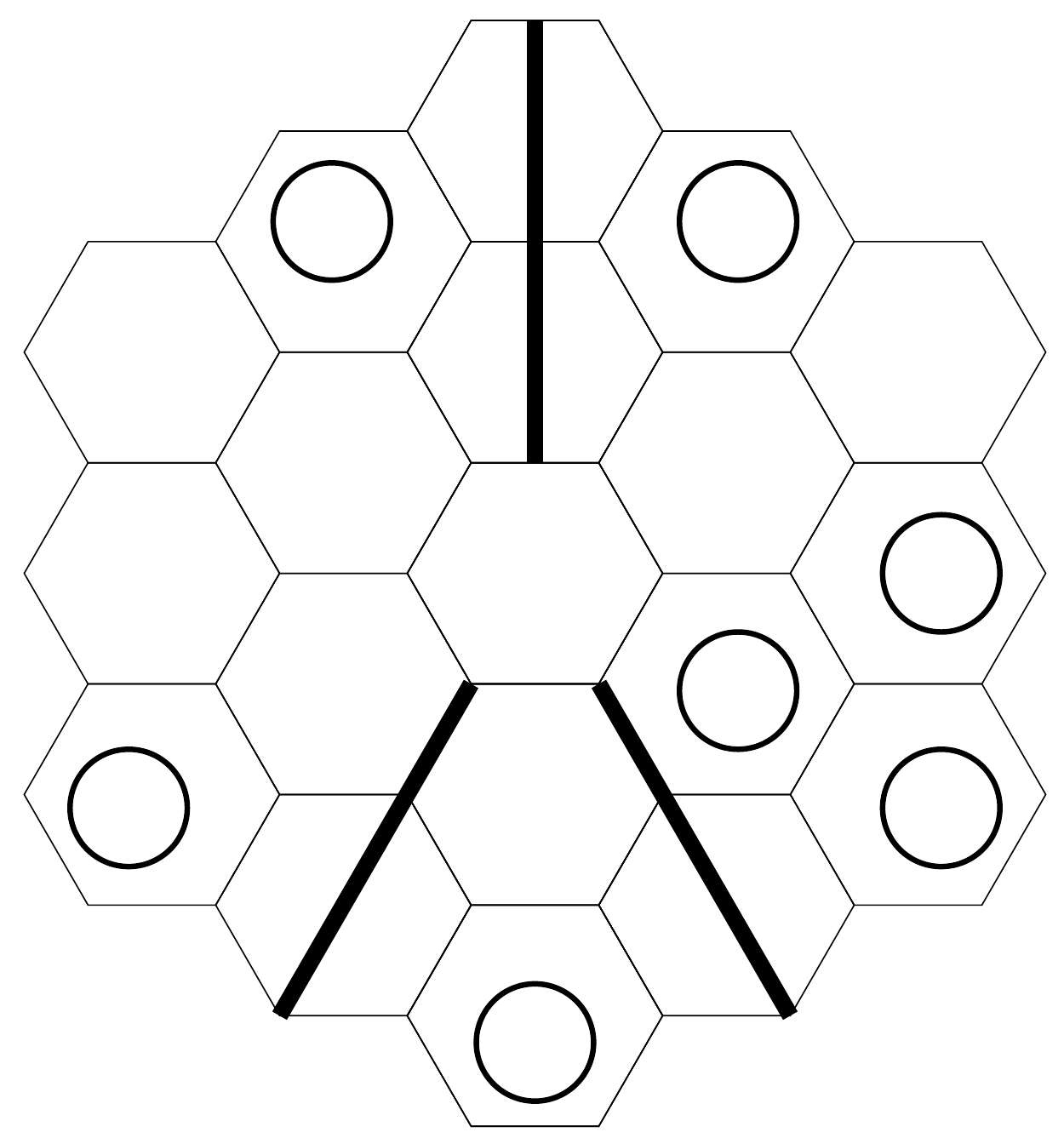} 
   \includegraphics[angle=0,scale=0.375]{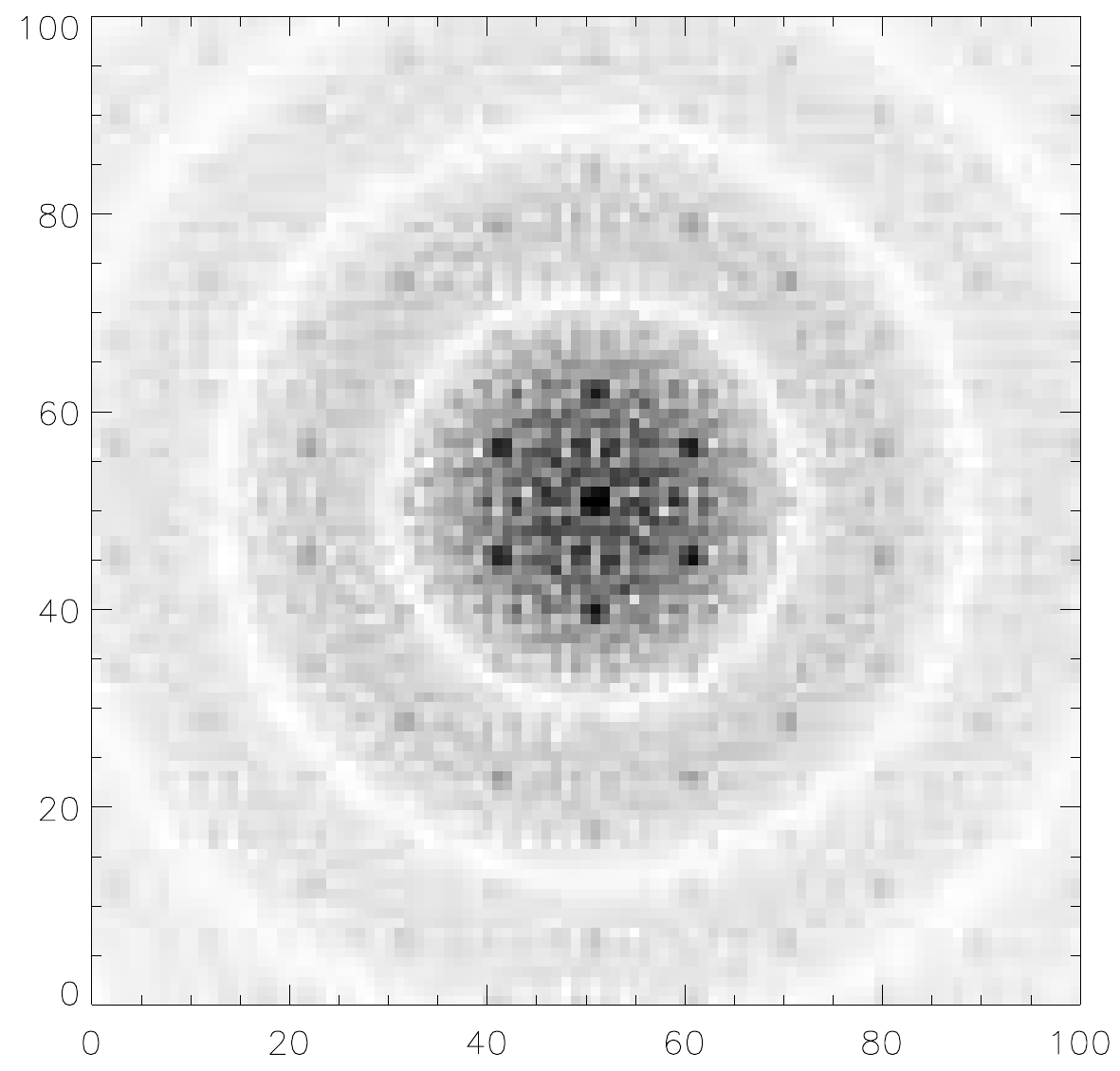}
   \includegraphics[angle=0,scale=0.375]{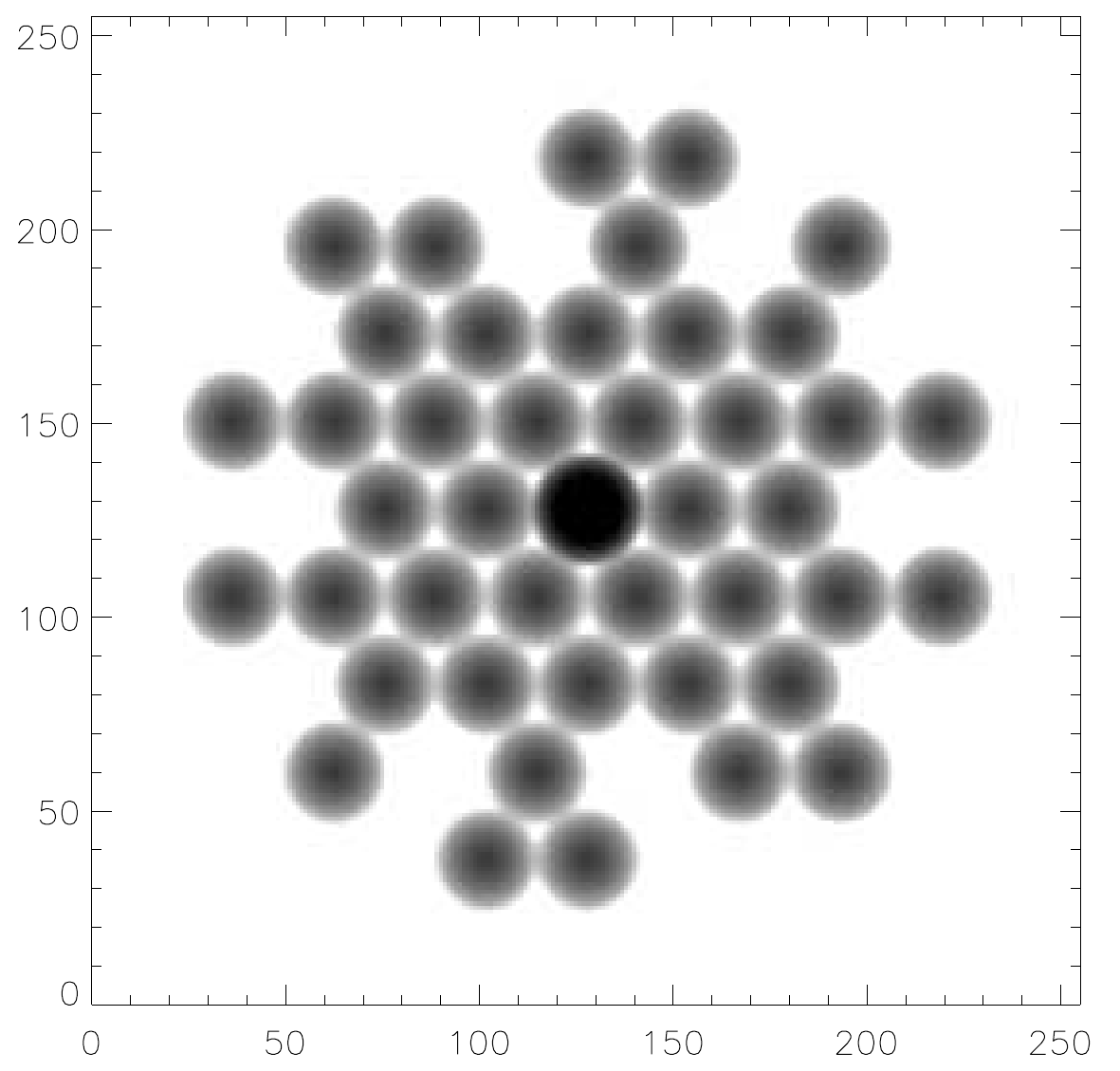} }
 \caption{
   {\it Left:} A 7-hole non-redundant pupil mask (holes) overlaid on
   the \JWST\  primary mirror.
   {\it Center:} A 4.05 \micron\  FGS-TFI image of a point source, with
   a 1\% bandwidth filter, on a negative logarithmic stretch (dark
   Airy rings appear white). The first Airy ring diameter is
   $2.4\arcsec$.  Generally data out to the first Airy ring is
   analyzed.
   {\it Right:} the power spectrum of the image (also on a negative
   log scale), showing fringe power on the 21\,baselines passed by the
   mask.}
 \label{mask7fig}
 \end{figure}
Interferometry with non-redundant baselines was intially developed for
radio astronomy \citep{1958MNRAS.118..276J}, and adapted
to optical wavelengths \citep{1986Natur.320..595B}. Today it is used
in IR and optical bandpasses \citep{1998SPIE.3350..839T}.
A detailed description of NRM can be found in
\citet{1989MNRAS.241P..51H,2000PASP..112..555T,2003RPPh...66..789M}
and references therein.
In brief, an N-hole pupil mask turns the extremely redundant
full aperture of a telescope into a simpler interferometric
array. Taking into account the geometry of the pupil (segmentation,
central obscuration and spider vanes) as well as the spectral bandwidth, the
mask is designed such that each baseline (a vector linking the centers
of two holes) is unique: the mask is non-redundant in the sense
that each spatial frequency is only sampled once (Fig.~\ref{mask7fig}).

In a sense, NRM is a form of \textit{imaging} with an unusual-looking
PSF that displays multiple sharp peaks, whereas a traditional
diffraction-limited PSFs is dominated by only one peak. However, the real
virtues of the NRM PSF only appear after the image is Fourier
transformed (e.g. Fig.~\ref{mask7fig} right panel): all spatial
frequencies sampled by the mask appear as well separated peaks
containing both amplitude and phase information. The advantages of
this approach are that:
\begin{itemize}\addtolength{\itemsep}{-0.5\baselineskip}
\item
The longest baseline provides a resolving power of $0.5\lambda/D$, 
compared to the traditionally accepted full aperture
Rayleigh criterion  limit of $1.22 \lambda/D$.
\item
Unlike conventional PSF co-addition, dominated by a speckle noise floor
\citep{2002ApJ...581L..59S, 2007ApJ...669..642S}, the {\bf information}
extracted from the NRM data {\bf can be averaged} to reduce noise,
even in the presence of slowly-varying speckles.
\item
Non-redundancy ensures that the complex visibilities can be used to
form {\bf closure phases} \citep{1986Natur.320..595B}, an observable
that {\bf calibrates} wavefront residual errors as well as {\bf
  non-common path errors} between the science and sensing arms of the
instrument.
\item
The outer working angle (OWA) is set by the shortest baseline,
typically $4 \lambda/D$: the {\bf NRM} search space nicely
{\bf complements} that of {\bf coronagraphy}.
\end{itemize}

It may seem perverse to throw away most of a telescope's collecting area,
but the 10~--~20\% mask transmission price paid for NRM (comparable to most high performance coronagraphs'
throughput) purchases not only
a straightforward $2.44\times$ gain in resolution but also
a dramatic increase in signal to noise ratio.
Today's ground-based NRM routinely achieves stability of 0.5 degrees
on the closure phase, hence passively stabilizing the phase at the
level of $\lambda/500$ to $\lambda/1000$, performance that will only
be matched by the next generation of ExAO coronagraphic systems
\citep{2008ApJ...688..701S}.

Still, ground-based NRM suffers from rapid temporal instabilities due to
atmospheric scintillation and transparency variations, as well as from
differential atmospheric refraction. Space will provide an exceptionally
stable environment that necessarily translates to increased 
performance.
Space-based NRM contrast limits will then set by detector and telescope details:
flat fielding accuracy, target placement repeatability, charge
diffusion, intrapixel sensitivity, charge transfer efficiency, 
details of guiding and pupil wander, and the drift of the telescope
structure due to slowly-varying thermal loads.

\begin{figure}
	\centerline{ \includegraphics[height=0.250\textwidth]{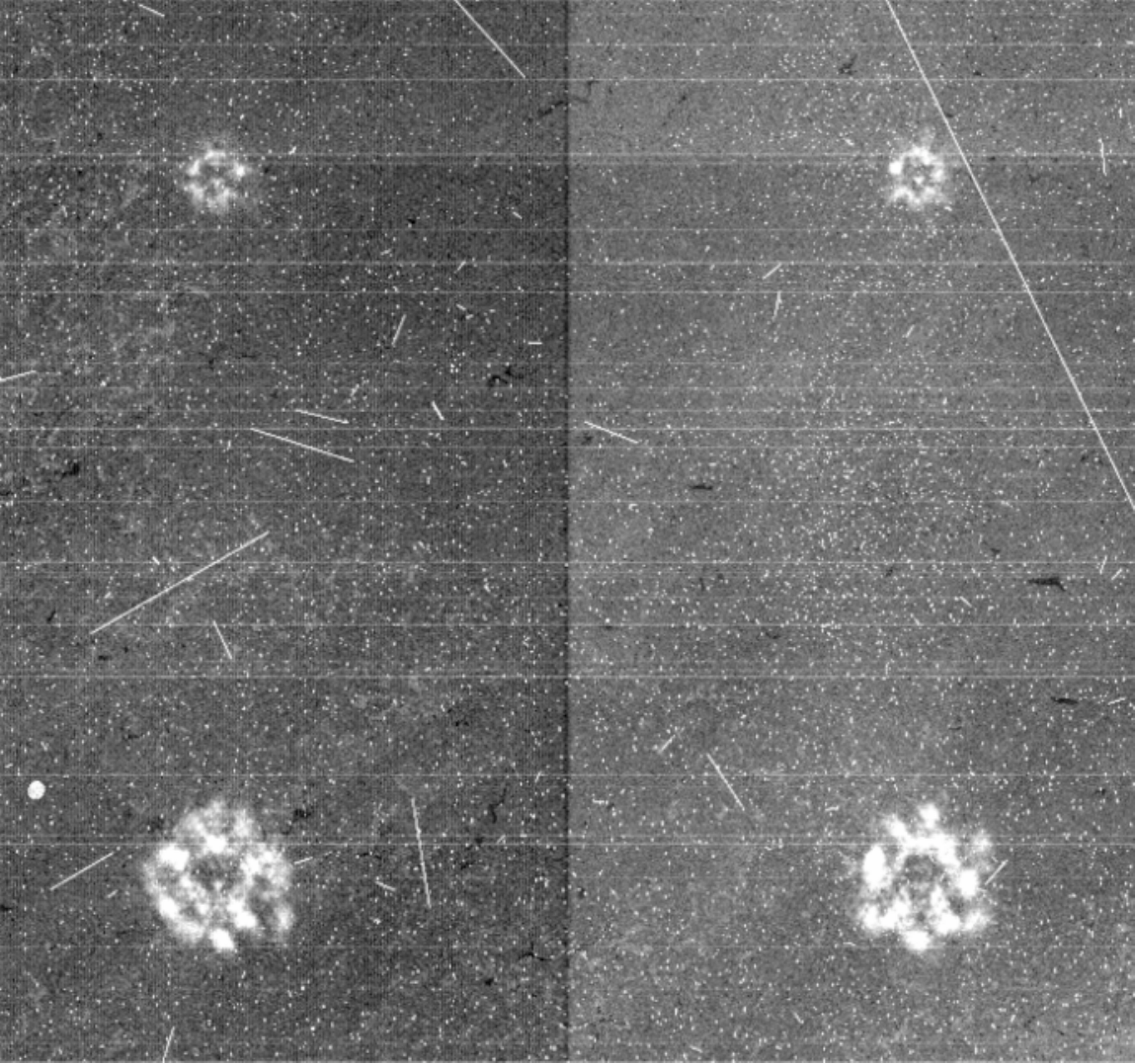} 
	             \includegraphics[height=0.250\textwidth]{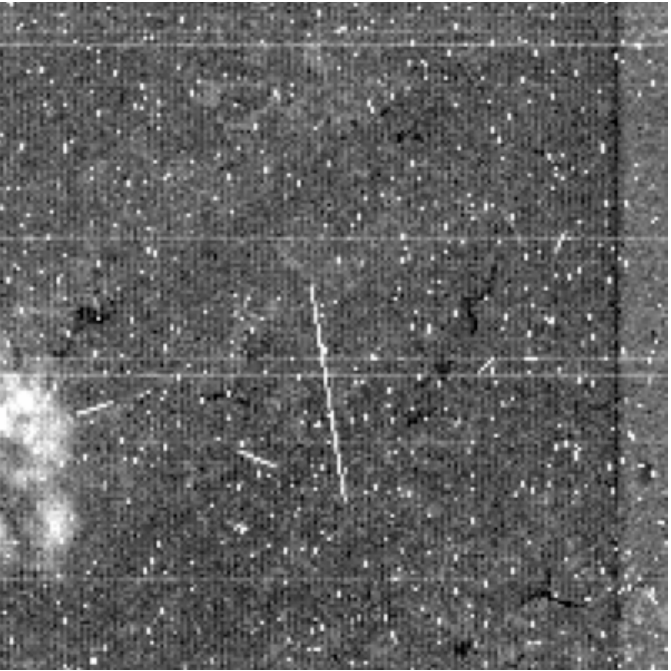} }
	\caption{  A JWST NIRCam HAWAII-2RG detector simulation using lab noise data
	\citep{Anand2003SPIE,2004SPIE.5487..909S}.
	\textit{Left:} full detector.  \textit{Right:} section of detector.
	Realistic NRM data can be modelled similarly on
	IR Focal Plane Arrays (FPA) and CCDs.  }
	\label{TwoImages}
\end{figure}

Accurate target placement on a detector can mitigate flat fielding
uncertainties. Requirements on target placement and their relation to
miscalibrated detector flat field responses must be derived from real
CCD and NIR FPA detector data. Intrapixel sensitivity variations and
charge transfer efficiency effects can be constrained with detector
simulations (e.g. \citealt{Anand2003SPIE,2004SPIE.5487..909S},
see Fig.~\ref{TwoImages}). 
Telescope stability and mechanical drift effect simulations already exist
\citep{Makidon08}, using ultrafine subpixel PSF generation 
\citep{2007OExpr..1515935S}.
Finite element analysis (FEA) modelling of JWST
\citep{2004SPIE.5487..825L} can incorporate 
breathing due to changes in the thermal environment of a segmented
telescope,  though monolithic 4 and  8~m  space telescope  should also be studied
in the NRM context.

NRM filter bandwidths up to 20\% are feasible. 
To utilize the performance capabilities of NRM fully 
it will be necessary to tackle two key tasks: 
first, the creation of realistic simulations of in-band/out-of-band spectral features
such as methane, or emission/absorption lines around AGNs and GBHCs
and, second, disk modeling for all types of astrophysical disks, with particular
attention to predicted polarization.  
Although not attempted yet, polarimetric observations are perfectly compatible with NRM.
The extra level of calibration added by polarimetry will translate into a higher
overall sensitivity of the technique. Such high angular resolution polarimetry
will require a sustained effort in disk modeling to develop strong science cases.

\section{Existing simulations and data analysis}

\citet{JAM09} utilized a sample set of ten independent realizations of JWST's
pupil within its wavefront error budget \citep{2004SPIE.5487..825L}
to estimate the impact of mechanical drift on NRM contrast.
JWST guiding errors and measured intra-pixel quantum efficiency were modeled
to create images on the FGS-TFI pixel scale.
Simulated data were analyzed with a pipeline used for Keck, Palomar, and VLT data.
(\eg\ \citealt{2008ApJ...679..762K}).
Nine of the ten PSF realizations in a set were taken to be calibrator observations,
and one of the ten was taken to be the target observation.
This process was repeated for several choices of target PSF.
A pessimistic 5$\sigma$ detection threshold was calculated, with closure-phase and
visibility standard deviations estimated from the scatter within the
9 statistically independent calibrator observations. 
Fig.~\ref{nrmmdpfig} (right panel) shows the contrast achieved with dynamic range calculated
two ways: (1) using only the closure phases (labelled CLP) for a comparison with
current ground-based results, and (2) using closure phases and fringe visibilities
simultaneously (labelled CLP+VIS),
to estimate the full range of contrast available to JWST. 
These methods need to be extended to cover the NIR to UV range, with adequately
fine pixel sizes and anisotropic guiding errors, to illuminate the science role of NRM
on proposed 4 to 16~m space telescopes.

\section{Optical metrology and mission risk mitigation}
  \label{sec:phasing}

In addition to its scientific merits,
NRM can provide optical system metrology data with interferometric accuracy.
A non-redundant array exhibits the remarkable property that each Fourier component 
in the image plane corresponds to a unique mapping to a pair of
subapertures in the pupil. This mapping is an ill-posed problem for a full aperture.
Direct metrology of the entire pupil can therefore be obtained with a
finite number of simple measurements using the science detector.

This is particularly advantageous to any telescope with a segmented
primary mirror. Every NRM image measures the relative mirror phase
between each hole in the aperture mask, with a capture range set by
filter coherence lengths; for a narrowband filter this can easily
be several hundred microns.
With measurements of the phase at multiple wavelengths
\citep{2004JOptA...6..216M,2005A&A...429..747B}, inter-segment piston
can be measured to nanometer precision, starting from an initially disordered state.
\textbf{No segment movement, defocus, or PSF degradation is required to
perform wavefront sensing anywhere in the field of view: a mask in the science camera pupil suffices}.

Similarly, any science which requires a thorough understanding of temporal, 
field-dependent or chromatic PSF variations will
benefit from precision metrology enabled by NRM. 
Thus NRM is a precise and flexible addition to the optical engineer's toolkit,
providing unique diagnostics and mitigating mission risk.


\newcommand{\etal}{{\it et al.} }
\newcommand{\asca}{{\it ASCA} }
\newcommand{\einstein} {{\it Einstein} } 
\newcommand{\exosat}{{\it EXOSAT} }
\newcommand{\ginga}{{\it Ginga} }
\newcommand{\bbxrt}{{\it BBXRT} }
\newcommand{\rosat}{{\it ROSAT} }
\newcommand{\iue}{{\it IUE} }
\newcommand{\fuse}{{\it FUSE} }
\newcommand{\xmm}{{\it XMM-Newton} }
\newcommand{\chandra}{{\it Chandra} }
\newcommand{\conx}{{\it Constellation-X} }
\newcommand{\hst}{{\it HST}/STIS }   
\newcommand{\rxte}{{\it RXTE} }
\newcommand{\hetg}{{\it HETGS} }
\newcommand{\letg}{{\it LETGS} }
\newcommand{\fekalfa}{{Fe~K$\alpha$} }
\newcommand{\bsax}{{\it BeppoSAX} }
\newcommand{\hi}{H~{\sc i} }        
\newcommand{\cfour}{C~{\sc iv} }
\newcommand{\csix}{C~{\sc vi} }
\newcommand{\nfive}{N~{\sc v} }
\newcommand{\sifour}{Si~{\sc iv} }
\newcommand{\cthree}{C~{\sc iii} }
\newcommand{\silya}{Si~{\sc xiv}~Ly$\alpha$ }
\newcommand{\nelya}{Ne~{\sc x}~Ly$\alpha$ }
\newcommand{\nlya}{N~{\sc vii}~Ly$\alpha$ }
\newcommand{\oxlc}{O~{\sc viii}~Ly$\gamma$ ($\lambda 15.176$) } 
\newcommand{\oxla}{O~{\sc viii}~Ly$\alpha$ }
\newcommand{\mglya}{Mg~{\sc xii}~Ly$\alpha$ }
\newcommand{\arla}{Ar~{\sc xviii}~Ly$\alpha$ ($\lambda 3.733$) }
\newcommand{\nala}{Na~{\sc xi}~Ly$\alpha$ ($\lambda 10.025$) }
\newcommand{\silb}{Si~{\sc xiv}~Ly$\beta$ ($\lambda 5.217$) }
\newcommand{\nelb}{Ne~{\sc x}~Ly$\beta$ ($\lambda 10.239$) }
\newcommand{\nilb}{N~{\sc vii}~Ly$\beta$ ($\lambda 20.910$) }   
\newcommand{\oxlb}{O~{\sc viii}~Ly$\beta$ ($\lambda 16.006$) }
\newcommand{\mglb}{Mg~{\sc xii}~Ly$\beta$ ($\lambda 7.106$) }
\newcommand{\oxythree}{O~{\sc iii} } 
\newcommand{\oxysix}{O~{\sc vi} } 
\newcommand{\oxyseven}{O~{\sc vii} }
\newcommand{\oxyeight}{O~{\sc viii} }
\newcommand{\oxysevenf}{[O~{\sc vii}] }
\newcommand{\oxyseveni}{O~{\sc vii} (i) }
\newcommand{\oxysevenr}{O~{\sc vii} (r) }
\newcommand{\nenine}{Ne~{\sc ix} }
\newcommand{\neninef}{[Ne~{\sc ix}] }
\newcommand{\neniner}{Ne~{\sc ix} (r) }
\newcommand{\neninei}{Ne~{\sc ix} (i) }
\newcommand{\neseven}{Ne~{\sc vii} }
\newcommand{\fetwenthree}{Fe~{\sc xxiii} }
\newcommand{\fethirteen}{Fe~{\sc xiii} }
\newcommand{\fefourteen}{Fe~{\sc xiv} }
\newcommand{\fetwelve}{Fe~{\sc xii} }
\newcommand{\fefifteen}{Fe~{\sc xv} }
\newcommand{\feseventeen}{Fe~{\sc xvii} }
\newcommand{\feeighteen}{Fe~{\sc xviii} }  
\newcommand{\fenineteen}{Fe~{\sc xix} }
\newcommand{\fetwenty}{Fe~{\sc xx} }
\newcommand{\fetwentyone}{Fe~{\sc xxi} }
\newcommand{\fetwentytwo}{Fe~{\sc xxii} }
\newcommand{\fetwentythree}{Fe~{\sc xxiii} }
\newcommand{\fetwentyfour}{Fe~{\sc xxiv} }
\newcommand{\fetwentyfive}{Fe~{\sc xxv} }
\newcommand{\fetwentysix}{Fe~{\sc xxvi} }
\newcommand{\felya}{Fe~Ly$\alpha$ }
\newcommand{\mgeleven}{Mg~{\sc xi} }
\newcommand{\mgelevenr}{Mg~{\sc xi} (r) }
\newcommand{\sinine}{Si~{\sc ix} }
\newcommand{\sithirteen}{Si~{\sc xiii} }
\newcommand{\sithirteenr}{Si~{\sc xiii} (r) }
\newcommand{\sulfsixteenf}{[S~{\sc xvi}] }
\newcommand{\res} {${\it 1s^{2}-1s np}$ }
\newcommand{\resonetwo} {${\it  1s^{2}-1s 2p}$ }
\newcommand{\resonethree} {${\it 1s^{2}-1s 3p}$ }
\newcommand{\resonefour} {${\it 1s^{2}-1s 4p}$ }
\newcommand{\src}{NGC~4593 }
\newcommand{\ic}{IC~4329A }
\newcommand{\thr}{3C~120 }
\newcommand{\ngc}{NGC~4593 }
\newcommand{\mkn}{Mkn~509 }
\newcommand{\mk}{Mkn~766 }
\newcommand{\mcg}{MCG~$-$6$-$30$-$15 } 
\newcommand{\ak}{Akn~564 }

\section{Planet and star formation science with NRM in space}
\label{sec:psfscience}

NRM's combination of very small inner working angle (IWA) with moderately
high contrast opens the door to unique studies of star and planet
formation inaccessible to other techniques. For example, the discovery
that the transition disk surrounding CoKu/Tau4 T Tauri star is
actually a circumbinary disk was made with NRM at Keck
\citep{2008ApJ...678L..59I}.
Here we present only a few of the most exciting science cases enabled by
space-based NRM.

An 8~m version of ATLAST with NRM would be sensitive to structure on
scales as small as 1.3 mas for 0.1 \micron, \textbf{almost an order of magnitude
finer than ALMA's finest resolution}.

\textsc{Extrasolar Planets and Circumstellar Disks}:
As mentioned above, aperture-masking on JWST will allow the detection
of young planets around nearby stars in the thermal infrared \citep{JAM09}.
There is justifiably great momentum for extending recent successes
in coronagraphic imaging of exoplanets to higher contrasts in order
to detect and characterize terrestrial planets.
But it is just as important to understand the formation process(es)
which lead to the observed diversity of exoplanets, a goal best
accomplished by observing ongoing
planet formation in protoplanetary disks on size scales similar to
our solar system (0.3-30 AU).

The closest such disks can be found in nearby star
forming regions such as Taurus and Ophiuchus at distances $d \sim
140$ pc.  JWST NRM can potentially detect young planets there with separations $> 12$ AU.
In Taurus, there are $\sim 60$ single stars with $M > 0.6~ M_\odot$; a survey of the
most promising 20 could require $\sim$ 50 hours (see Fig.~\ref{nrmmdpfig}).
NRM contrast should not degrade much for binary targets (unlike coronagraphy), so in
theory NRM can probe a larger population of planets by targeting binary systems as well. However,
further development of data analysis techniques for circumbinary planets
is required before a major JWST survey is undertaken.

For future space telescopes, the improvement in IWA versus conventional
coronagraphy brings within reach both previously unobservable regions
of nearby systems as well as an increased survey space for more
distant targets: NRM is able to observe targets in Orion with spatial
resolution comparable to that achieved by coronagraphy for systems in
Taurus (albeit at a lower contrast, but still one sufficient to detect
luminous hot young giant planets). In conjunction with an 8-16 m space
telescope, NRM may enable us to directly investigate how the
properties of planetary systems vary with formation environment
between loosely distributed T associations and the more massive OB
associations.

Visible and near-UV observations with NRM on a $\geq 8$ m
space telescope offer the potential to detect structures such
as gaps or spiral density waves on scales of $0.2-5$ AU within disks
in Taurus.  Such observations directly probe the zones where
terrestrial and Jovian planets are built, and will allow us to
disentangle the relative contributions of planet formation models such
as gravitational instability and core accretion
\citep{astro2010Kraus}.
Furthermore, they will let us observe planet/disk
interactions to verify models for angular momentum transport and
orbital migration.

Multiwavelength NRM imaging and polarimetry can be used to further
constrain both disk structure and the nature of scattering bodies
\citep{Watson:2007p2642}.  For outer regions of disks accessible to
existing coronagraphs (at separations $> 30$ AU), reflectance spectroscopy has been
used to infer the presence of icy mantles on dust grains
\citep[e.g.][]{Honda:2009p2692} and possibly even organic chemicals
\citep{2008ApJ...673L.191D}.  NRM will, for the first time, extend
such observations inward to the region of the ``snow line'' predicted
around 4-5 AU (Fig.~\ref{nrmmdpfig}).  Measurements of water ice absorption bands at 1.5,
2.0, and 3.1 \micron\  will verify predictions of how particle
properties vary across the snow line, directly relevant for planet
formation models that posit buildup of planetesimals at the snow line
\citep{2007ApJ...664L..55K}.


\textsc{Bipolar Outflows} are launched from disks
in
astrophysical contexts ranging from protoplanetary
systems to supermassive black holes \citep[e.g.][]{2007prpl.conf..231R}.
Several theories have been advanced to explain their origin, but the physical mechanism(s) responsible for acceleration and
collimation remain unclear.
Discriminating between the various models
requires spatially resolving the launch region
\citep{2007prpl.conf..231R,2006AnA...453..785F}, but the best
observations to date (from
HST STIS and AO integral field spectroscopy; \citealt{2008LNP...742..151B}
and references therein) are limited
to resolutions of $\sim 15$ AU for the closest jets,
compared to an inferred launch
region size of $\sim 5$ AU \citep{2004ApJ...609..261H}.

NRM on an 8-m space telescope will be able to fully resolve the
launching regions of jets from nearby YSOs, with a spatial resolution of $\sim 1$ AU at 656 nm for $d=140$ pc.
By observing jet opening angles and
collimation scales, these observations will enable us to directly
discriminate between disk and X-wind models. Such observations could
be obtained using 1\% wide filters, but NRM would be even more powerful
in conjunction with a moderate-resolution optical integral
field spectrograph, which would provide invaluable kinematic information
as well as improved contrast by reducing the background of
disk and star light against which jet emission lines must be detected.
YSO jets in Taurus are the closest examples of collimated
astrophysical outflows, and thus offer the best laboratory for
detailed, high angular resolution study of the same physics which
underlies jet production in systems as diverse as X-ray binaries and
AGN.

\section{Black hole environs science with NRM in space}
\label{sec:blackholescience}

Accretion flows around black holes (BH) cannot yet be imaged directly, rather we 
infer structure and composition from spectral information (e.g. 
\citealt{2009MNRAS.394..491M} \& references therein).  
For supermassive BH (SBH~masses $\sim 10^{6}-10^{9} M_{\odot}$), NRM will 
offer a direct view of critical parts of the accretion flow, providing a 
window on the nature and rate of mass supply to AGN, the orientation of the 
accretion flow around AGN, the accretion environment
 of low luminosity AGN (LLAGN) and will even permit a look at the kinematics 
of material as close as 80 AU to Sgr A*, the closest SBH.
For Galactic black hole candidates (GBHC~masses of  few--few hundred $M_{\odot}$)
space-based NRM will image their jets and accretion flows in unprecedented 
detail (0.6mas at 0.1$\micron$ or lengthscales of $\sim 1.2$AU for Cyg X1).

\textsc{Supermassive black holes:} SBH live in the centers of galaxies, and can
 power AGN. Observed AGN properties
 are believed to depend on the orientation of the 
accretion flow to the observer's sightline \citep{1993ARA&A..31..473A} and its
 obscuration, with Type 1 AGN face-on and Type 2 edge-on to the observer. 
Space-based NRM imaging at 0.6-24mas together with polarimetry can provide unprecedented information 
on the orientation of the accretion disk, the disk outer structure (clumpy or 
smooth) and the AGN-host galaxy connection (for appropriate systems). NRM 
imaging resolution will be an order of magnitude finer 
than expected with ALMA ($\sim$5-34mas at 0.5--3.6mm in the most extended 18km 
baseline configuration) \citep{2008NewAR..52..339M} or in preliminary 
coronagraphic studies of AGN 
($\sim$ 70mas) \citep{2005A&A...429..433G}. With the addition of spectral
information from a TFI or IFU we can map the kinematics of the accretion flow. A 
JWST-TFI, has a kinematic resolution of $\sim 600$km$\rm{s}^{-1}$ at $\sim 5\micron$,
which will allow us to map the velocity field of the accretion disk and
the Broad Line Region in Type 1 AGN. 

NGC 1068 is the 
prototypical Seyfert 2 (Type 2) AGN, at a distance of $12.6$Mpc with $M_{sbh} 
\sim 10^{7} M_{\odot}$  and a dusty torus $\sim 1.7$pc ($\sim 17$mas) in 
radius \citep{2004Natur.429...47J}. Space-based NRM imaging at 0.6-24mas can 
probe the structure of the torus (internal heat sources, clumpiness), the 
narrow line region (NLR) clouds as well as the link between the outer edge of the torus and the host galaxy. Since the AGN phenomenon is believed 
to scale with SBH mass (e.g. \citealt{2009MNRAS.394..491M} \& references therein), an SBH $100$ times more massive than that in NGC 1068 (i.e.~$10^{9} M_{\odot}$) could be probed at the same structural resolution as 
NGC 1068, at 10 times the distance ($\sim 126$Mpc). Lower luminosity AGN 
(LLAGN and LINERs) may be either highly obscured SBH or low accretion rate 
SBH. At 
$\sim 16$Mpc, NGC 3718 is classified as both a Seyfert 1 and a LINER with a 
central $M_{BH} \sim$ few $\times 10^{7} M_{\odot}$ and could be probed by 
space-based NRM (with a 16m baseline) on scales of $\sim 0.05-2$pc, which 
spans the outer edge of the accretion disk and most of the dusty torus. In our own Galaxy, 
Sgr A* is an SBH ($\sim 3 \times 10^{6} M_{\odot}$) with Schwarzschild radius 
of $\sim 0.1$AU \citep{2000MNRAS.317..348G} or $\sim 0.013$mas. 
At $\sim 2 \micron$ NRM can probe to within $\sim$80~AU of Sgr A* and 
monitor the motion of mass around a quiescent SBH on unprecedented scales. 

\textsc{Black hole growth, AGN evolution and cosmic merger history of galaxies:}
SBH mass seems to correlate with bulge
 luminosity and stellar velocity dispersion (i.e. elliptical galaxies and S0 
galaxies tend to host larger mass SBH) \citep{1995ARA&A..33..581K,1998AJ....115.2285M,2000ApJ...539L...9F,2000ApJ...543L...5G,2005ApJ...619L.151B,2006ApJ...641L..21G}. If SBHs 
formed from the merger of earlier (possibly dwarf) galaxies, then the 
orientation of the accreting system should be random in large bulges and 
correlated with the host spiral galaxy orientation in galaxies with little or 
no bulge. NRM imaging combined with polarimetry will map AGN orientation 
with great accuracy, allowing us to test models of SBH growth and the host 
galaxy-SBH relation.

\textsc{Stellar mass black holes:} GBHCs are the least luminous black holes, but because they are nearby, they 
are relatively easy to detect, usually as X-ray bright companions in X-ray 
binaries. The changing `states' of GBHCs are believed to be caused by changes in the mass accretion rate, which triggers significant changes in the accretion flow geometry \citep{2001ApJS..132..377H}. Cyg X1, 
the original GBHC, is relatively typical at $\sim 2$kpc distance and is 
separated from its companion by $\sim 0.2$AU ($\sim 0.1$mas), with a period of
 $\sim 5.6$ 
days and a modulation period of $\sim 142$ days, possibly caused by accretion 
disk precession \citep{1999MNRAS.309.1063B}. Cyg X1 has an intermittent relativistic 
radio-jet with an opening angle $<2^{o}$ extending $\sim 15$mas from the 
core \citep{2001MNRAS.327.1273S}. At 0.1-4$ \micron$, NRM resolution 
is 0.6-24mas with a 16~m baseline which will reveal spectacular details in 
jet evolution and accretion outflows from close to the black hole.
GBHC provide a laboratory to study the 
evolution of jets on timescales 
$\sim 10^{-6}$ faster than in AGN \citep{2002Sci...298..196C}. 

\section {Conclusion}
Non-redundant aperture masking has established itself as a powerful and productive high
angular resolution technique, and is implemented on most major ground
based observatories.  It was recently accepted as a late addition to JWST's
FGS-TFI.
There are good reasons: detailed simulations with time-varying mirror
figure errors and existing data reduction methods suggest
that non-redundant aperture masks would benefit any of the JWST
instruments, bringing exciting high resolution high contrast imaging
within reach.
In fact, low cost, minimal impact on hardware, relaxed optical
requirements, and high scientific payoff recommend NRM as a leading
technology for any future missions possessing a high angular
resolution imaging component.

\section {Recommendations}
\begin{itemize}\addtolength{\itemsep}{-0.5\baselineskip}
\item We recommend the development of a coronagraphy-NRM instrument
	for proposed 4, 8, and 16~m space telescopes at the Conceptual Design Review level.
	Scientific and technical trades between IFUs, tunable filters, or a filter wheel
	as spectral discriminators for NRM must be developed, with downselect at the end of 5 years.
	Detailed detector effects and telescope FEA results must be taken into account, and 
	NRM data analysis methods developed further.
	[Cost: 5M\$, Time: 5 years].
	This should be followed by systematic development of the chosen NRM design(s) [5M\$, 5 years].
\item An exhaustive list of space-based NRM science opportunities, 
	with exposure times and stability and hardware requirements,
	needs to be developed.  Polarization must be an integral part
	of this effort. [1M\$, 5 years].
\item Experimental verification and development of NRM, like other novel techniques,
	should be supported on existing ground-based telescopes and 
	laboratory testbeds. [5M\$, 10 years].
\item Early NRM work on Keck and Palomar provided a platform for the development
	of the ideas expounded here. Further ground-based observations with NRM should be
	supported to develop techniques and enhance the science return from future space-based NRM
	instruments and missions.  [5M\$, 10 years].

\item NRM is capable of sensing mirror phase to interferometric precision.
	We recommend its study and testing as a versatile, robust wavefront sensing system
	for future space telescopes, starting immediately with existing testbeds.
	[10M\$, 5 years].  
\end{itemize}

\begin{multicols}{2}
{
\scriptsize
\bibliographystyle{apj-short}
\bibliography{ms}
}
\end{multicols}

\end{document}